\documentstyle[11pt]{article}
\begin{document}

\title{{\bf Phase-Mixing and Dissipation of Standing Shear Alfv\'{e}n waves}}
\author{K. Karami$^{1,2}$,\thanks{E-mail: KKarami@uok.ac.ir}\\
Z. Ebrahimi${^1}$\\$^{1}$\small{Department of Physics, University
of Kurdistan, Pasdaran Street, Sanandaj,
Iran}\\$^{2}$\small{Research Institute for Astronomy $\&$
Astrophysics of Maragha (RIAAM), Maragha, Iran}}

\maketitle

\begin{abstract}
We study the phase mixing and dissipation of a packet of standing
shear Alfv\'{e}n waves localized in a region with non-uniform
Alfv\'{e}n background velocity. We investigate the validity of the
exponential damping law in time, $\exp(-At^3)$, presented by
Heyvaerts \& Priest (1983) for different ranges of Lundquist, $S$,
and Reynolds, $R$, numbers. Our numerical results shows that it is
valid for $(R,S)\geq 10^7$.
\end{abstract}
\clearpage
\section{Introduction}
%Please see the PASA Style Guide for help with correct layout for your manuscript.
%Examples of tables and figures are given below.
\label{intro}

Since the discovery of the hot solar corona by Edl\'{e}n (1943),
different theories of coronal heating have been put forward and
debated. Heyvaerts \& Priest (1983), hereafter HP83, were first to
suggest that the phase-mixing of Alfv\'{e}n waves in coronal
plasmas could be a primary mechanism in coronal heating. They
showed that the phase-mixing occurs due to inhomogeneity of the
local Alfv\'{e}n phase speed across the background magnetic field.
HP83 analytically showed that in both the strong phase-mixing
limit and the weak damping approximation, the amplitude of
standing Alfv\'{e}n waves decays with time as $\exp(-At^3)$ where
$A$ is a function of the coordinate corresponding to the
inhomogeneous direction ($x$, in this paper). Since then, much
analytical and numerical work has been done on the subject. Nocera
et al. (1984) studied the phase mixing of propagating Alfv\'{e}n
waves in an inhomogeneous medium. They pointed out if transverse
gradients are smeared out as soon as they are formed, this yields
to weak phase mixing where damping laws differ from solutions of
HP83. Parker (1991) investigated the effect of a density and/or
temperature gradient in the direction of vibration of a transverse
Alfv\'{e}n wave. The result was a strong coupling of the waves on
different lines of force, producing a coordinated mode that was
not subject to simple phase mixing. Hood et al.(1997a) derived a
self similar solution of the Alfv\'{e}n wave phase-mixing
equations for heating of coronal holes. They showed that the
damping of the waves with height follows the scaling predicted by
HP83 at low heights, before switching to an algebraic decay at
large heights. Hood et al. (1997b) obtained a simple, self similar
solution for the heating of coronal loops by phase mixing. They
showed the HP83 model still does work well in a certain class of
coronal loops and the phase mixing can supply heating at large
Lundquist number at timescales shorter than or comparable with the
radiative cooling timescale. Nakariakov et al. (1997) considered
the nonlinear excitation of fast magnetosonic waves by phase
mixing Alfv\'{e}n waves in a cold plasma with a smooth
inhomogeneity of density across a uniform magnetic field. They
suggested this nonlinear process as a possible mechanism of
indirect plasma heating by phase mixing through the excitation of
fast waves. But Botha et al. (2000) showed that the nonlinear
generation of fast modes by Alfv\'{e}n waves has little effect on
classical phase mixing. De Moortel et al. (1999) elaborated the
effect of density stratification on phase-mixing. They remarked
that when the inhomogeneity in the horizontal direction in the
plasma is sufficiently large, so the phase mixing is strong,
stratification is unimportant. In the other words due to the rapid
phase mixing, energy can be dissipated before the effects of
stratification build up. De Moortel et al. (2000) studied the
combined effect of a gravitationally stratified density and a
radially diverging background magnetic field on phase mixing of
Alfv\'{e}n waves. They found that: i) The efficiency of phase
mixing depends strongly on the particular geometry of the
configuration. ii) Depending on the value of the scale height the
wave amplitudes can damp either slower or faster than in the
uniform non-diverging model.

Hood et al. (2002) showed that the amplitude of single pulse and
bipolar pulse traveling in the z direction, contrary to infinite
wavetrain, have slower algebraic damping of the form $t^{-3/2}$
and $t^{-3}$, respectively, rather than exponential in time.
Tsiklauri et al. (2003) cleared that the decay rate of the
Alfv\'{e}nic part of a compressible 3D MHD pulse is affected
linearly by the degree of localization of the pulse in the
homogeneous transverse direction, but the dynamic of Alfv\'{e}n
waves can still be obtained from the previous 2.5D models, e.g.,
Hood et al. (2002). Smith et al. (2007) found that in presence of
the both density stratification and magnetic field divergence, the
enhanced phase mixing mechanism can dissipate Alfv\'{e}n waves at
heights less than half that was predicted by the previous
analytical solutions. They stated that if phase-mixing takes place
in strongly divergent magnetic fields, it is not necessary to
invoke anomalous viscosity in corona.

In the present work, we study the phase mixing of a packet of
standing shear Alfv\'{e}n waves in presence of the both viscous
and resistive dissipations. To do this, we numerically solve the
linearized MHD equations and obtain the damping time of the
oscillations. Our aim is to test the validity of HP83's damping
law for different ranges of the Reynolds and Lundquist numbers.
This paper is organized as follows. In Sect. 2, we introduce the
basic equations of motion and introduce the model. In Sect. 3, the
numerical results are reported, while the conclusions are given in
Sect. 4.

%-----------------------------------------------------------------------------------------------
\section{Equations of motion}
The linearized MHD equations for a zero-$\beta$ plasma are:
\begin{eqnarray} \frac{\partial\delta\textbf{v}}{\partial
t}=\frac{1}{4\pi\rho_{0}}\{(\nabla\times\delta\textbf{B})\times{\textbf{B}_{0}}
+(\nabla\times{\textbf{B}_{0}})\times\delta\textbf{B}\}
+\frac{\eta}{\rho_{0}}\nabla^2\delta\textbf{v},\label{mhd1}
\end{eqnarray}
\begin{eqnarray}
\frac{\partial\delta\textbf{B}}{\partial
t}=\nabla\times(\delta\textbf{v}\times{\textbf{B}_{0}})+
\frac{c^2}{4\pi\sigma}\nabla^2\delta\textbf{B},\label{mhd2}
\end{eqnarray}
where $\delta\textbf{v}$ and $\delta\textbf{B}$ are the Eulerian
perturbations in the velocity and magnetic fields; $\rho_{0}$,
$\sigma$, $\eta$ and $c$ are the mass density, the electrical
conductivity, the viscosity and the speed of light, respectively.

The simplifying assumptions are:
\begin{itemize}
      \item under coronal conditions gas pressure is negligible
(zero-$\beta$);
\item the equilibrium density profile is $\rho_{0}=\rho_{0}(x)$;
      \item there is a constant magnetic field along the $z$ axis, $\textbf{B}_{0}=B_{0}\mathbf{\hat{z}}$;
\item there is no initial steady flow inside or outside of the tube;
\item the viscous and resistive coefficients, $\eta$ and
$\sigma$ respectively, are constants.
   \end{itemize}

To solve Eqs. (\ref{mhd1})-(\ref{mhd2}), following HP83, we
neglect the variations in $y$ direction, $\frac{\partial}{\partial
y}=0$, and will further assume that the velocity perturbs in $y$
direction. So we choose a solution for the velocity perturbation
as
\begin{eqnarray}
\delta\textbf{v}(x,z,t)=\delta v_y(x,t)\sin(k z){\bf
\hat{y}},~~~~~~k=n\pi/L,\label{v-pert}
\end{eqnarray}
where $L$ is the length of the loop and $n=(1,2,3,...)$ is the
wave number in $z$ direction, respectively. Here the waves are
standing because of boundary conditions
$\delta\textbf{v}(x,0,t)=\delta\textbf{v}(x,L,t)=0$. Note that
HP83 also supposed that the loop is bounded from above an below by
boundaries at altitude $z=0$ and $z=L$. It is convenient to work
with dimensionless variables $\bar{x}=x/a$, $\bar{z}=z/a$,
$\bar{t}=t/\tau_{\rm A}$,
$\bar{\rho}_{0}(x)=\rho_{0}(x)/\rho_{00}$,
$\delta\bar{\textbf{v}}=\delta\textbf{v}/v_{\rm A_0}$ and
$\delta\bar{\textbf{B}}=\delta\textbf{B}/B_{0}$. Where $a$ is a
typical length scale of density inhomogeneity across the field
(i.e. loop radius) and $\tau_{A}=a/v_{\rm A_0}$ is  a time scale
for an Alfv\'{e}n wave to propagate along the inhomogeneity
direction. $\rho_{00}$ and $v_{\rm A_0}$ are the plasma density
and Alfv\'{e}n speed at $x=0$, respectively. Finally, Eqs.
(\ref{mhd1}) and (\ref{mhd2}) in dimensionless form, dropping the
'bars' for convenience, become
\begin{eqnarray}
\frac{\partial\delta v_y}{\partial
t}=v_{A}^2(x)\frac{\partial\delta B_y}{\partial
z}+\frac{1}{R}\nabla^2\delta v_y,\label{dimless1}
\end{eqnarray}
\begin{eqnarray}
\frac{\partial\delta B_y}{\partial t}=\frac{\partial\delta
v_y}{\partial z}+\frac{1}{S}\nabla^2\delta B_y,\label{dimless2}
\end{eqnarray}
where $v_{\rm A}(x)=\frac{1}{\sqrt{\rho_{0}(x)/\rho_{00}}}$ is the
dimensionless form of the Alfv\'{e}n speed. Also the Reynolds
number,
$$R=\Big(\frac{a^2\rho_{00}}{\eta}\Big)\Big/\Big(\frac{a}{v_{A_0}}\Big),$$
is the ratio of the viscous time-scale to the Alfv\'{e}n crossing
time, and the Lundquist number,
$$S=\Big(\frac{4\pi\sigma a^2}{c^2}\Big)\Big/\Big(\frac{a}{v_{A_0}}\Big),$$
is the ratio of the resistive time-scale to the Alfv\'{e}n
crossing time. Removing $\delta B_y$ from Eqs. (\ref{dimless1}) to
(\ref{dimless2}), and keeping only the first order-terms in $1/R$
and $1/S$ gives
\begin{eqnarray}
\frac{\partial^2\delta v_y}{\partial t^2}+k^2v_{\rm A}^2(x)\delta
v_y=\Big(\frac{1}{R}+\frac{1}{S}\Big)\nabla^2\frac{\partial\delta
v_y}{\partial t}+T(x,t),\label{main}
\end{eqnarray}
where
\begin{eqnarray}
T(x,t)=\frac{1}{S}\Big[6\Big(\frac{v_{\rm A}'}{v_{\rm
A}}\Big)^2-2\Big(\frac{v_{\rm A}''}{v_{\rm
A}}\Big)-4\Big(\frac{v_{\rm A}'}{v_{\rm
A}}\Big)\frac{\partial}{\partial x}\Big]\frac{\partial\delta
v_y}{\partial t},\label{T}
\end{eqnarray}
where prime indicates a derivative with respect to $x$. It is
obvious that the term $T(x,t)$ in Eq. (\ref{main}) becomes
important for high magnetic diffusion plasmas (low S) and in the
regions where the Alfv\'{e}n speed has a large gradient. For more
accuracy, in contrast with HP83, we keep this term in our
numerical simulations.

Since the dissipation rate is a function of $x$, for calculating
the overall damping time, it is suitable to calculate the
dimensionless total energy (kinetic energy plus magnetic energy)
of the packet per unit of length in the $y$ direction as
\begin{eqnarray}
\bar{E}_{\rm tot}(t)=\int_{0}^{2}\Big[\rho_{0}(x)\delta
v_y^2(x,t)+\delta B_y^2(x,t)\Big]dx,\label{Energy2}
\end{eqnarray}
where $\bar{E}_{\rm tot}(t)=\frac{16\pi}{B_0^2aL} E_{\rm tot}(t)$
and $\delta B_y(x,t)$ is calculated from Eq. (\ref{dimless2}).

\begin{figure}
\includegraphics{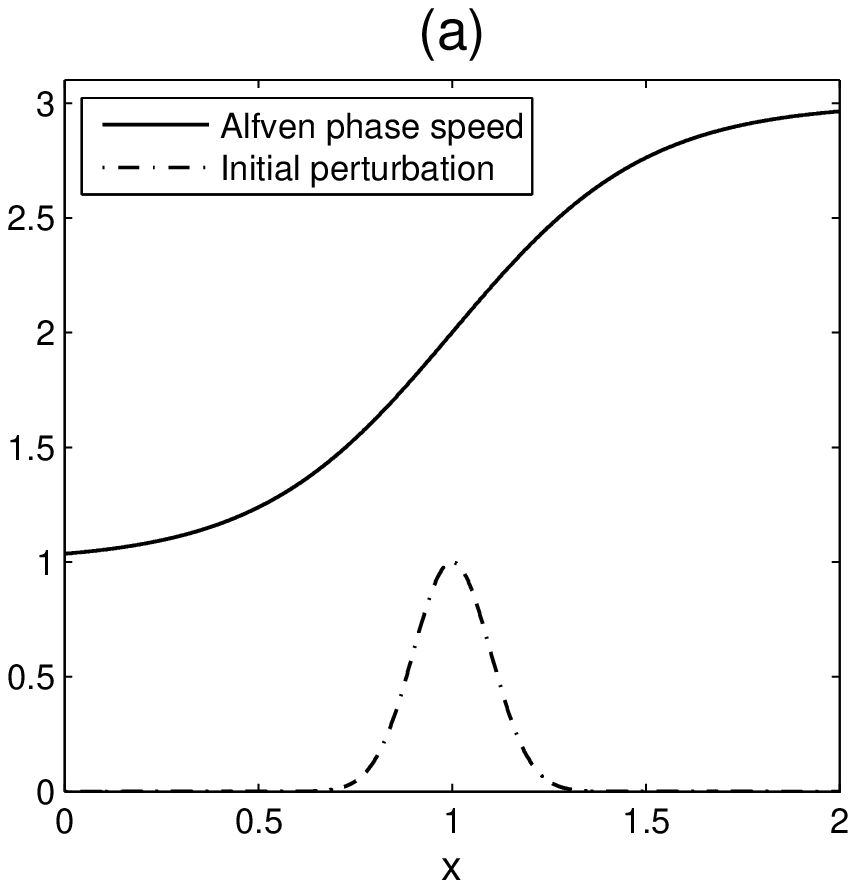}\includegraphics{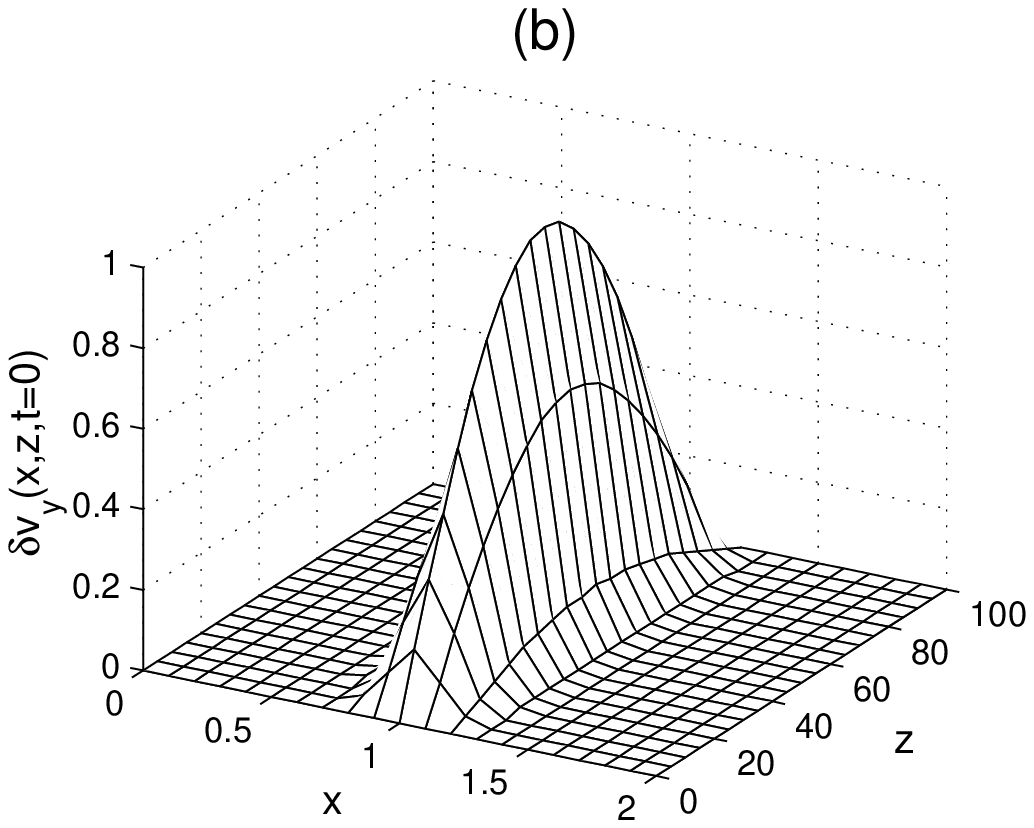} \vspace{3 cm} \caption[]{(a) The profile
of background Alfv\'{e}n speed (solid curve) and the initial
amplitude of velocity perturbations (dash-dotted curve) at $z=50$
as functions of x. (b) 3D view of the packet of standing
Alfv\'{e}nic pulses in fundamental mode ($n=1$). Auxiliary
parameters are: $\alpha=2$, $d=0.1$, and $L=100a$.} \label{set0}
\end{figure}

We suppose a functional form of dimensionless Alfv\'{e}n speed and
a Gaussian form of a localized packet of standing Alfv\'{e}n waves
around $x=1$ as
\begin{eqnarray}
v_{\rm A}(x)=2+\tanh[\alpha(x-1)],\label{V}
\end{eqnarray}
\begin{eqnarray}
\delta
v(x,z,t=0)=\exp{\Big[-\frac{1}{2}\Big(\frac{x-1}{d}\Big)^2\Big]}\sin(kz),\label{packet}
\end{eqnarray}
where parameter $\alpha$ controls the size of inhomogeneity and
$d$ is width of the packet. For $\alpha=2$ and $d=0.1$, the
Alfv\'{e}n speed profile and shape of the initial wave packet
given by Eqs. (\ref{V}) and (\ref{packet}) are plotted in Fig.
\ref{set0}, respectively.

Substituting Eq. (\ref{V}) in $\omega(x)=kv_{\rm A}(x)$ gives the
dimensionless average period of oscillation as
\begin{eqnarray}
P_{\rm
avg}^{\alpha}=\frac{1}{2}\int_{0}^{2}P(x)dx=\frac{\pi}{k}\int_{0}^{2}\frac{dx}{2+\tanh[\alpha(x-1)]}.\label{Pav}
\end{eqnarray}
%-----------------------------------------------------------------------------------------------
\section{Numerical Results}
As typical parameters for a coronal loop, we assume $L=10^5~{\rm
km}$, $a=10^3~{\rm km}$, $B_{0}=100{~\rm G}$, and
$\rho_{00}=2\times 10^{-14}~{\rm gr~cm^{-3}}$. For such a loop,
one finds $v_{\rm A_0}=2000~{\rm km~s^{-1}}$. Here the loop
parameters coincide with the TRACE observations (see Aschwanden et
al. 2002; Verwichte et al. 2004). We use a finite difference
method to solve Eq. (\ref{main}), numerically. The evolution of a
packet of fundamental standing Alfv\'{e}n modes is calculated in
the range of $0\leq x \leq 2$. To include the dynamical effect of
the exterior region, we let the wave packet to evolve up to $x=2$.
We suppose that the wave packet never reach at the $x$ boundaries.
Hence to avoid any contamination of the solution by the change of
boundary values, we fix the boundary conditions. This restricts
the time of simulation, but it is still possible to reach the
strong phase mixing limit. We choose the boundary and initial
conditions as
\begin{eqnarray}
\delta v_y(x=0,t)=\delta v_y(x=2,t)=0,\label{bound}
\end{eqnarray}
\begin{eqnarray}
\delta v_y(x,t=0)={\rm e}^{-50(x-1)^2},\label{init1}
\end{eqnarray}
\begin{eqnarray}
\frac{\partial\delta v_y(x,t)}{\partial t}\Big|_{t=0}=0.
\end{eqnarray}

There is an upper limit for the time of simulation because we can
simulate the evolution until any excitement near the $x$
boundaries could be occurred. The truncation error of numerical
results is $\Delta=O(\Delta t^3)+O(\Delta x^4)$. We should be
aware of choosing suitable spatial step size $\Delta x$, because
in the limit of strong phase mixing, large gradients in the $x$
direction are made, so the smaller $\Delta x$ is needed.

\begin{figure}
\includegraphics{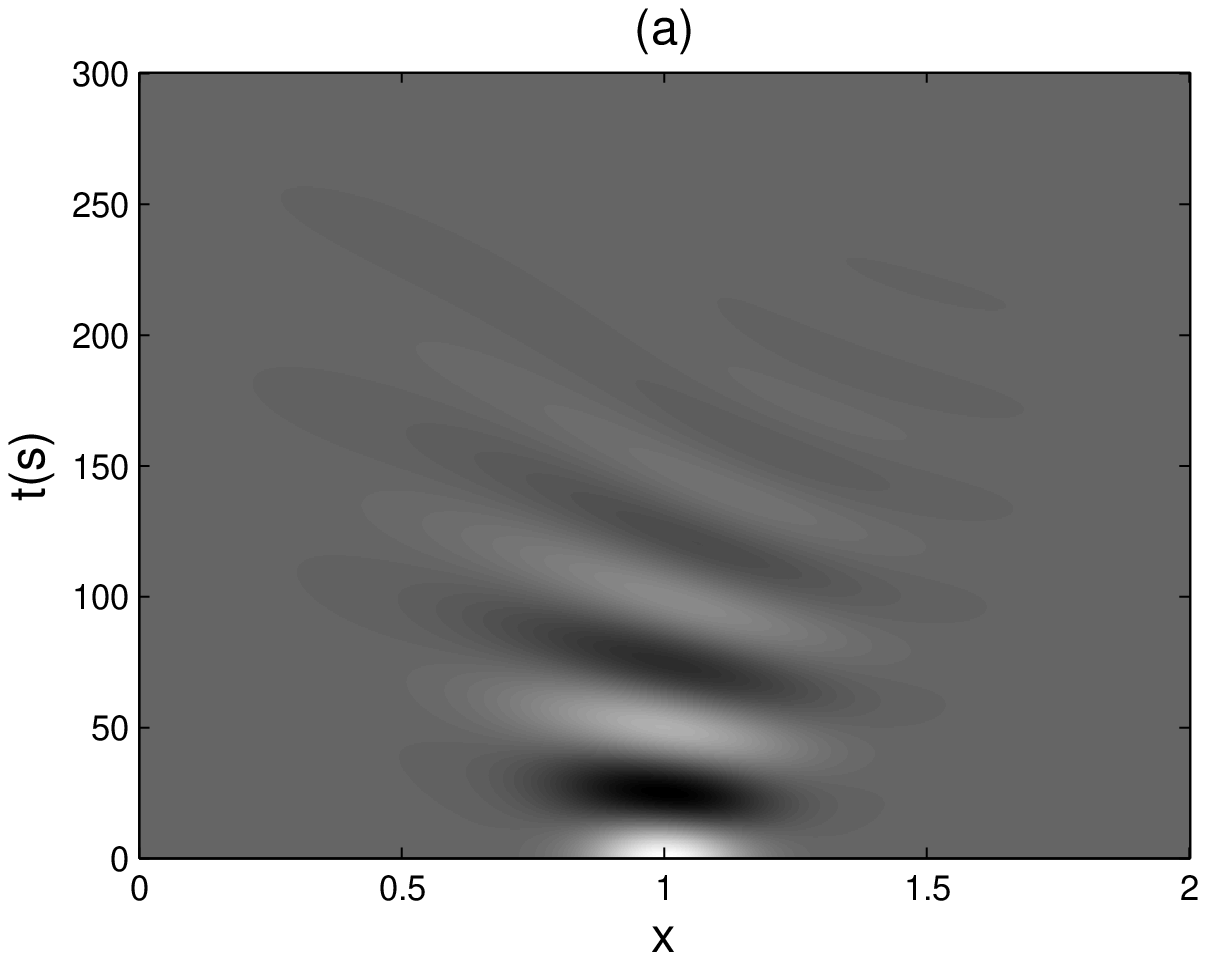}\includegraphics{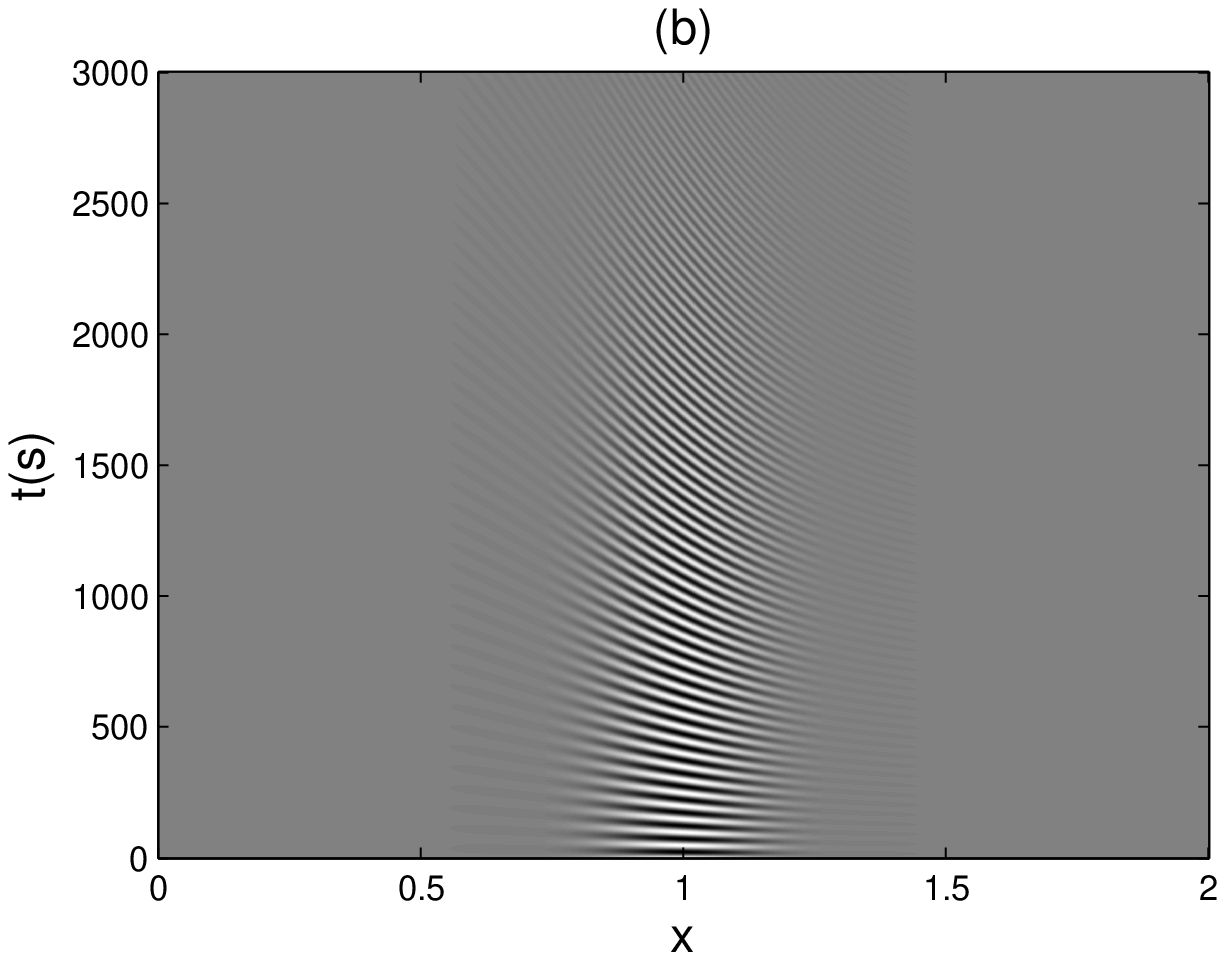} \vspace{0.0 cm} \caption[]{The contour
plot of $\delta v_{y}(x,t)$ in the case (a) $R=S=10^4$ ; (b)
$R=S=10^8$ for $d=0.1$ and $\alpha=2$.} \label{contour}
\end{figure}

\begin{figure}
\includegraphics{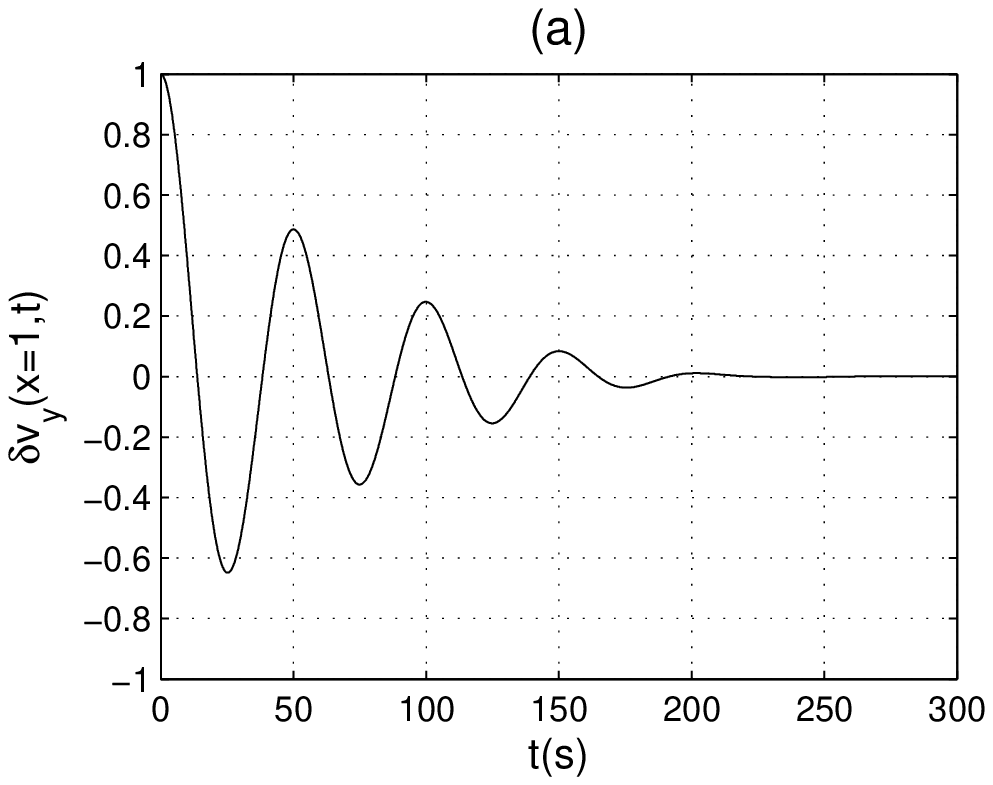} \includegraphics{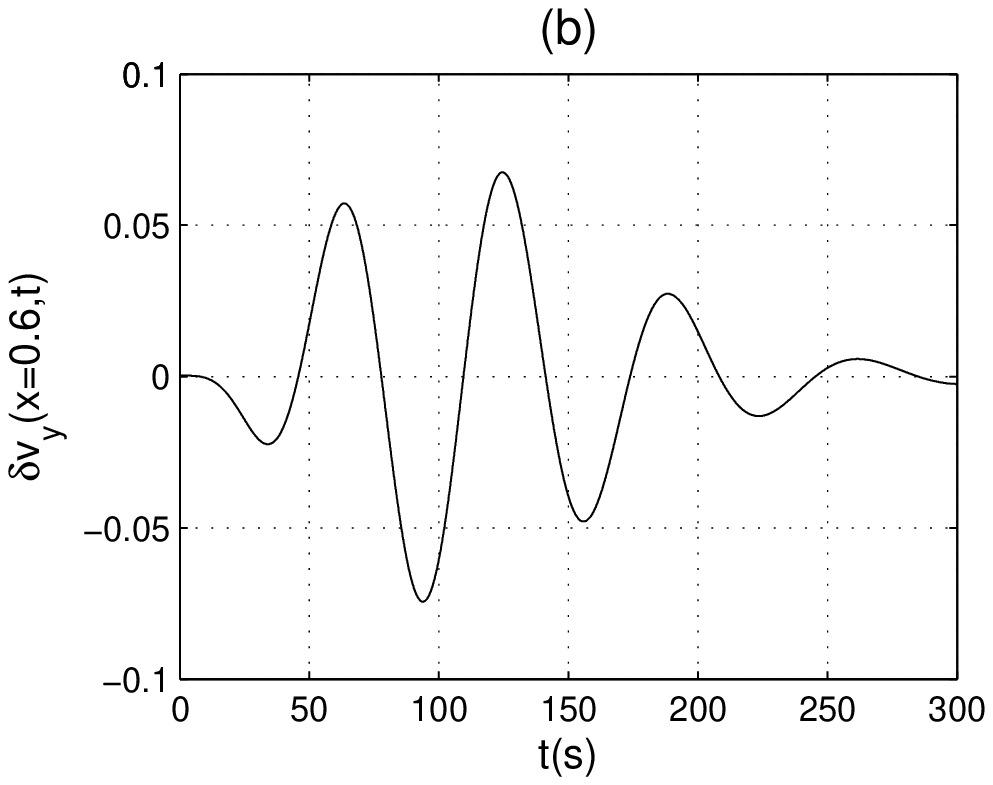} \includegraphics{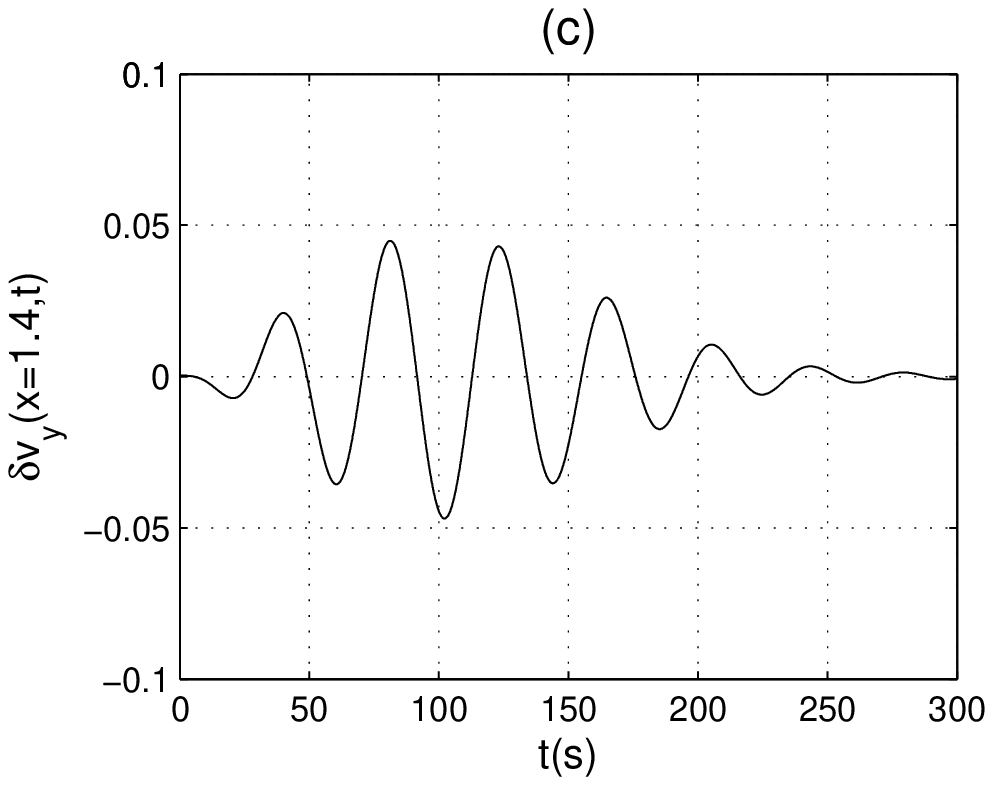} \vspace{2 cm} \caption[]{Cross
section cuts of $\delta v_y$ along (a) center of the packet
($x=1$); (b) $x=0.6$ (in the lower Alfv\'{e}n speed region); (c)
$x=1.4$ (in the higher Alfv\'{e}n speed region) for $R=S=10^4$,
$\alpha=2$ and $d=0.1$.} \label{setA-v}
\end{figure}

\begin{figure}
\includegraphics{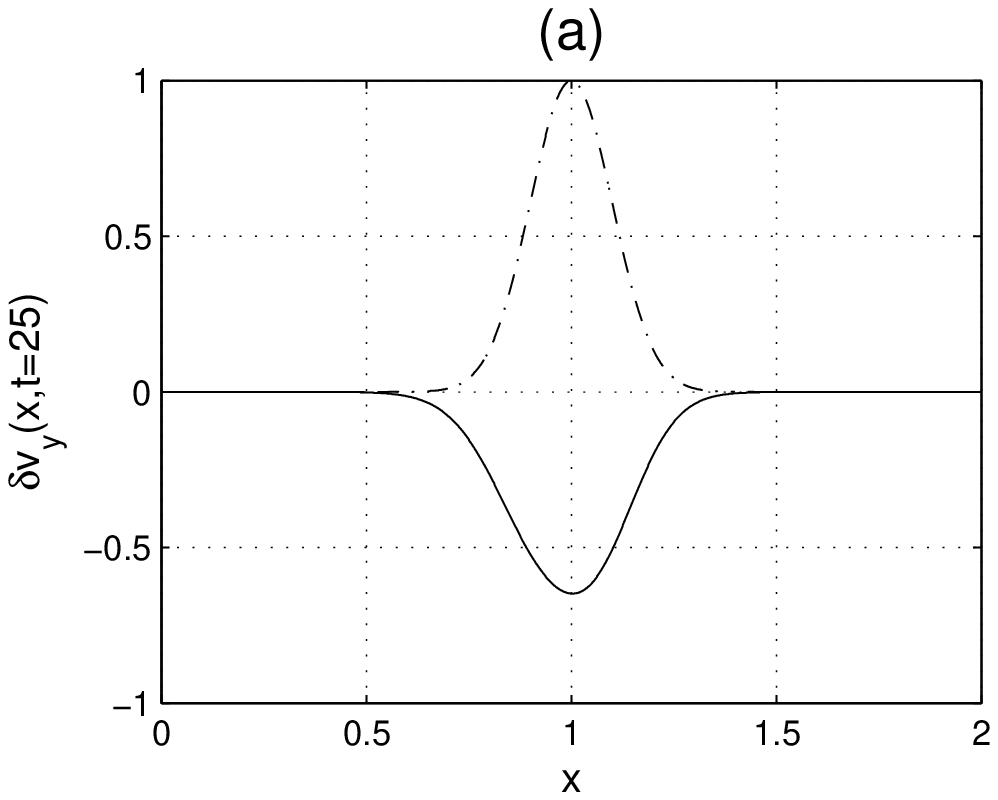} \includegraphics{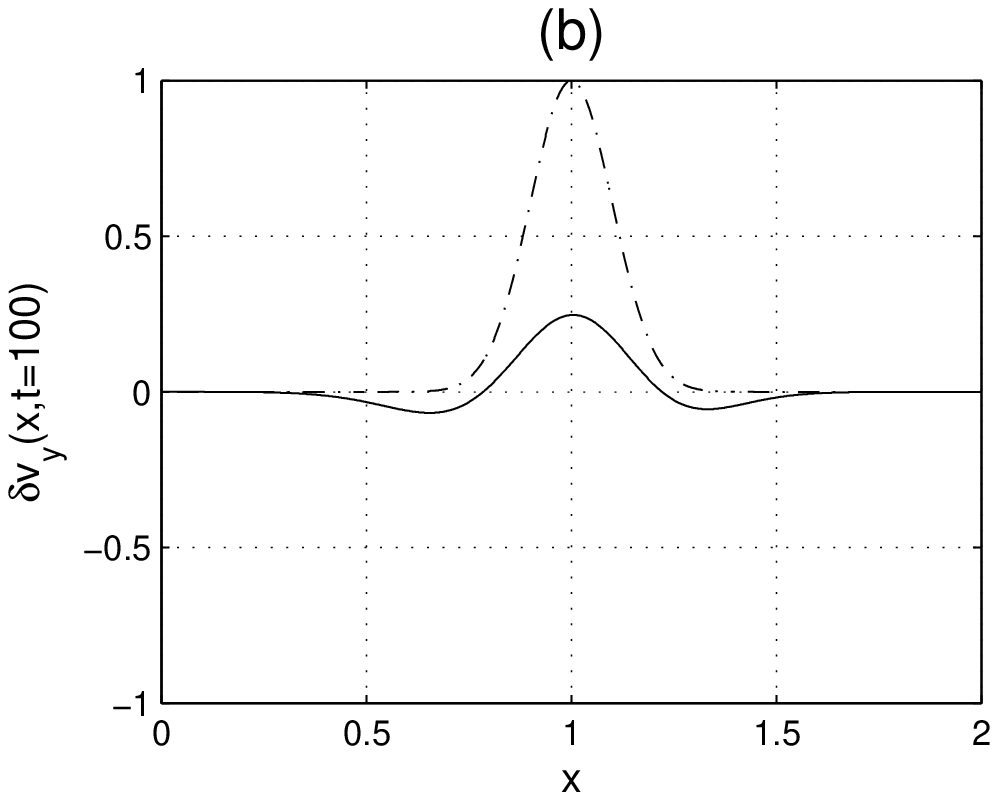} \includegraphics{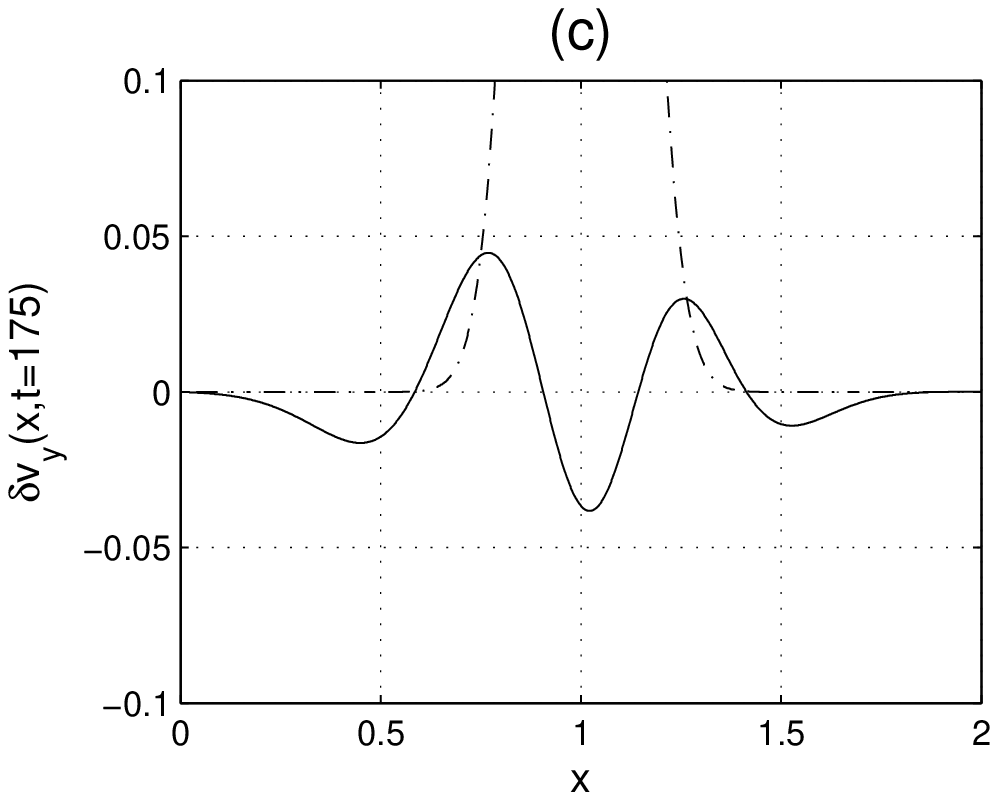} \vspace{2 cm} \caption[]{The
shape of the packet at (a) $t=25 s$ ; (b) $t=100 s$ ; (c) $t=175
s$ for $R=S=10^4$, $\alpha=2$ and $d=0.1$. The dash-dotted curve
is the initial gaussian packet.} \label{setA-t}
\end{figure}

\begin{figure}
\includegraphics{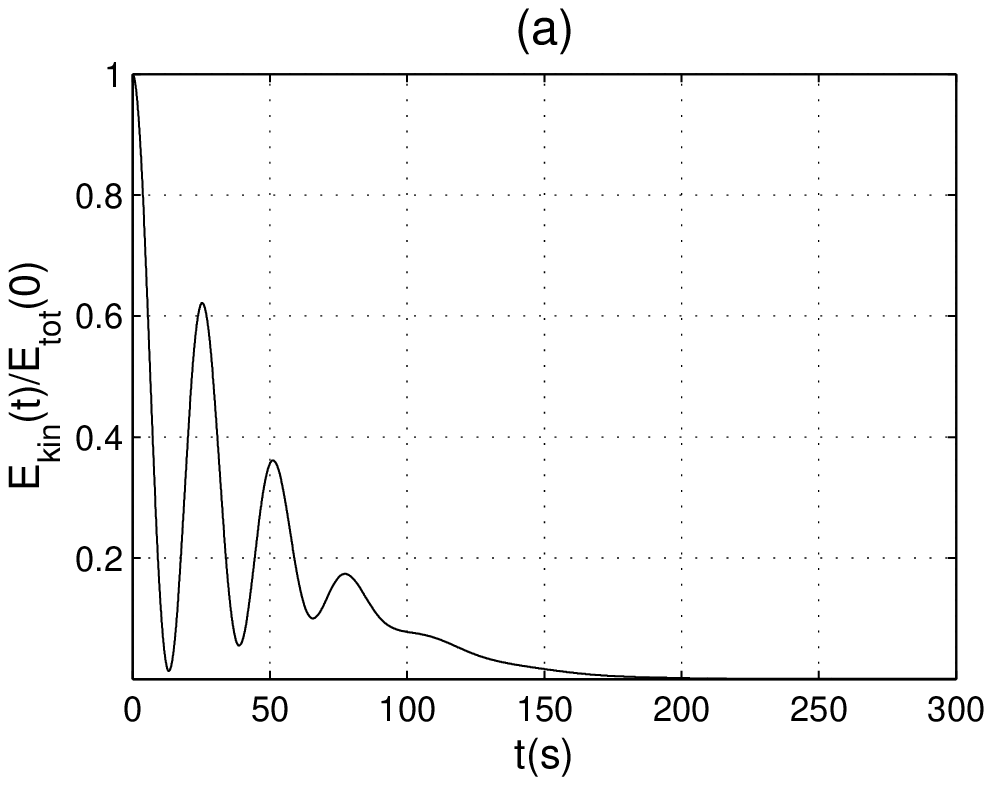} \includegraphics{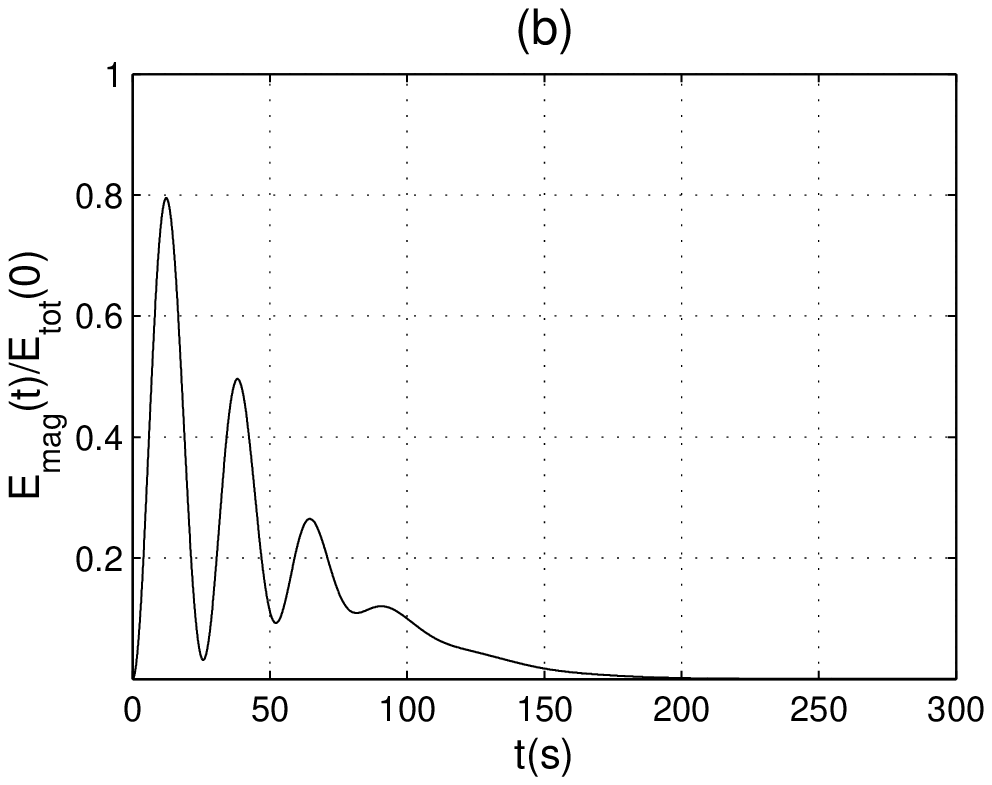} \includegraphics{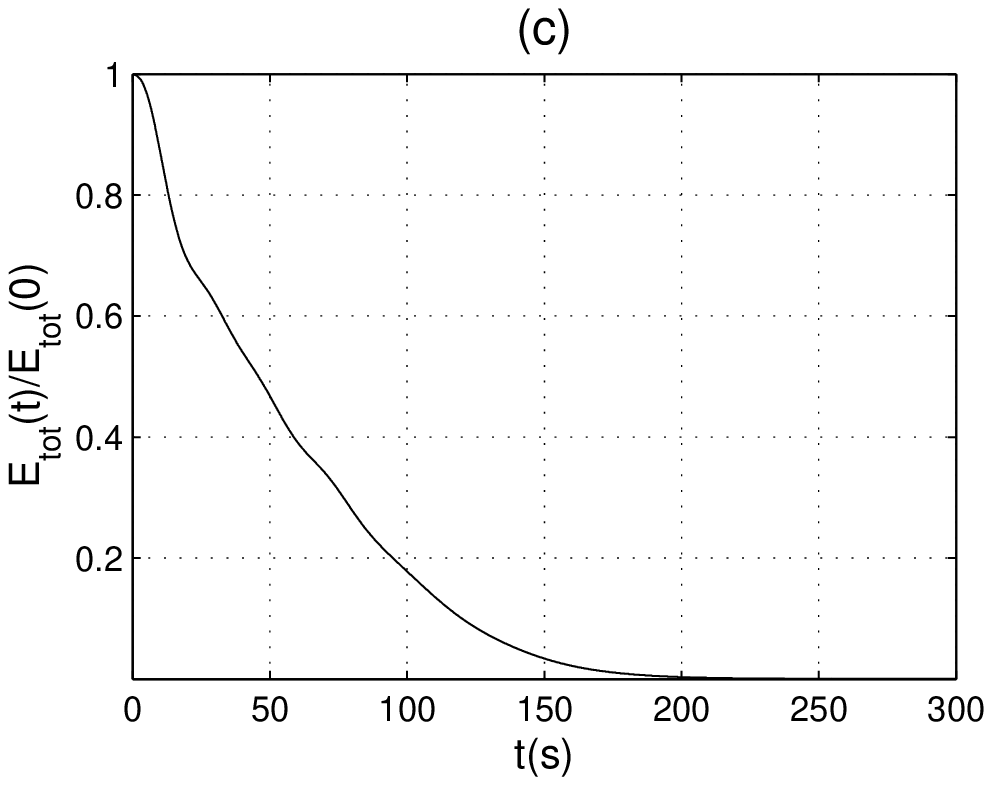} \vspace{2 cm} \caption[]{The
time dependencies of overall (a) kinetic energy; (b) magnetic
energy and (c) the decay rate of total energy of the wave packet
for $R=S=10^4$, $\alpha=2$ and $d=0.1$.} \label{setA-energy}
\end{figure}

Fig. $\ref{contour}$ shows contour plots of $\delta v_y(x,t)$ in
the $x-t$ plane for two different cases with  $R=S=10^4$ (a) and
$R=S=10^8$ (b). The white and black colors represent positive and
negative values of $\delta v_y(x,t)$, respectively. Fig.
$\ref{contour}$ clears that the defocusing of the packet in the
case (a) is large but not in the case (b). This is because of
coupling of oscillations in neighboring field lines due to
presence of damping terms in the right hand side of Eq.
(\ref{main}). Fig. \ref{setA-v} presents the cross-section cuts
along $x=1$, $x=0.6$ and $x=1.4$ for the case (a) with $\alpha=2$
and $d=0.1$. It illustrates that as central regions of the packet
decay with time, the neighboring oscillations in the regions with
smaller amplitudes, are excited and finally are damped by phase
mixing. This means that the packet defocuses along the x direction
which is illustrated in Fig. \ref{setA-t}. Fig. \ref{setA-energy}
shows the time evolution of the kinetic energy, magnetic energy
and total energy of the packet. Fig. \ref{setA-energy} reveals
that both the kinetic and magnetic energies of the packet
oscillate with time sharply at initial stage of the evolution and
then smoothly damped.

\begin{figure}
\includegraphics{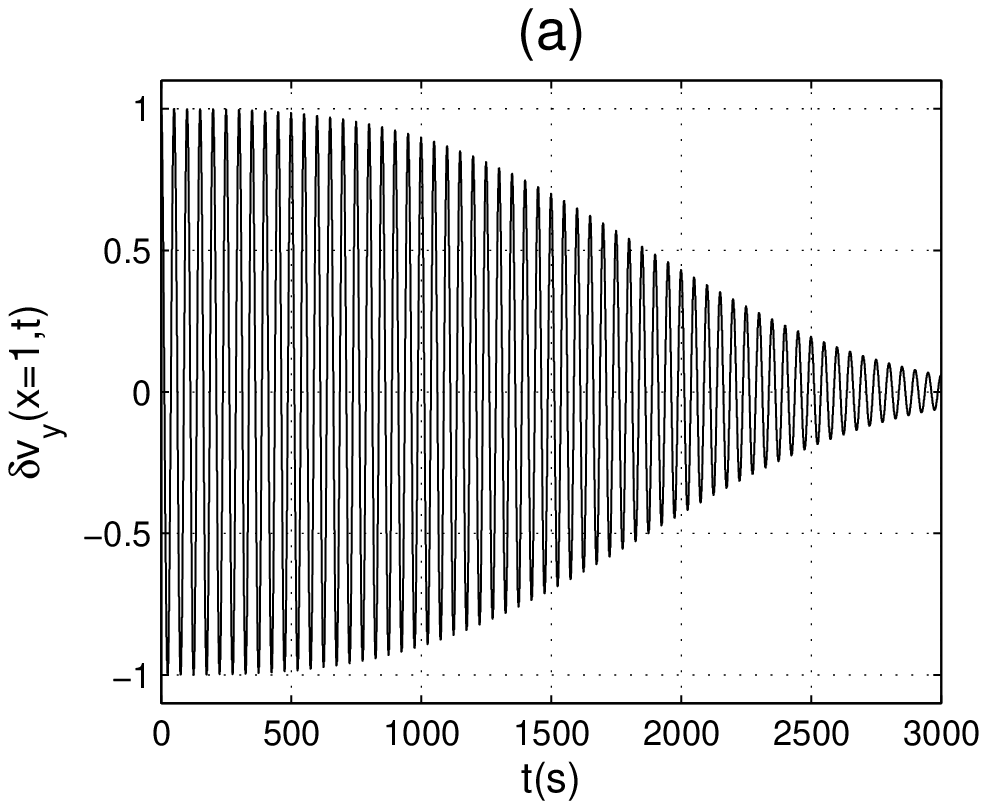} \includegraphics{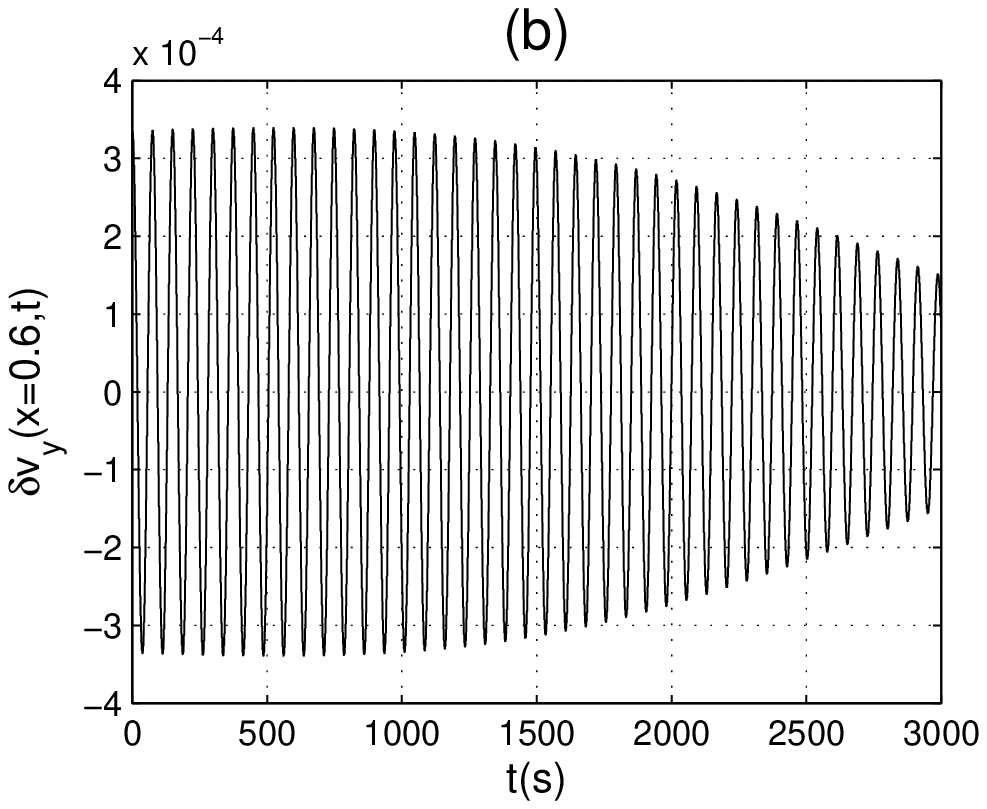} \includegraphics{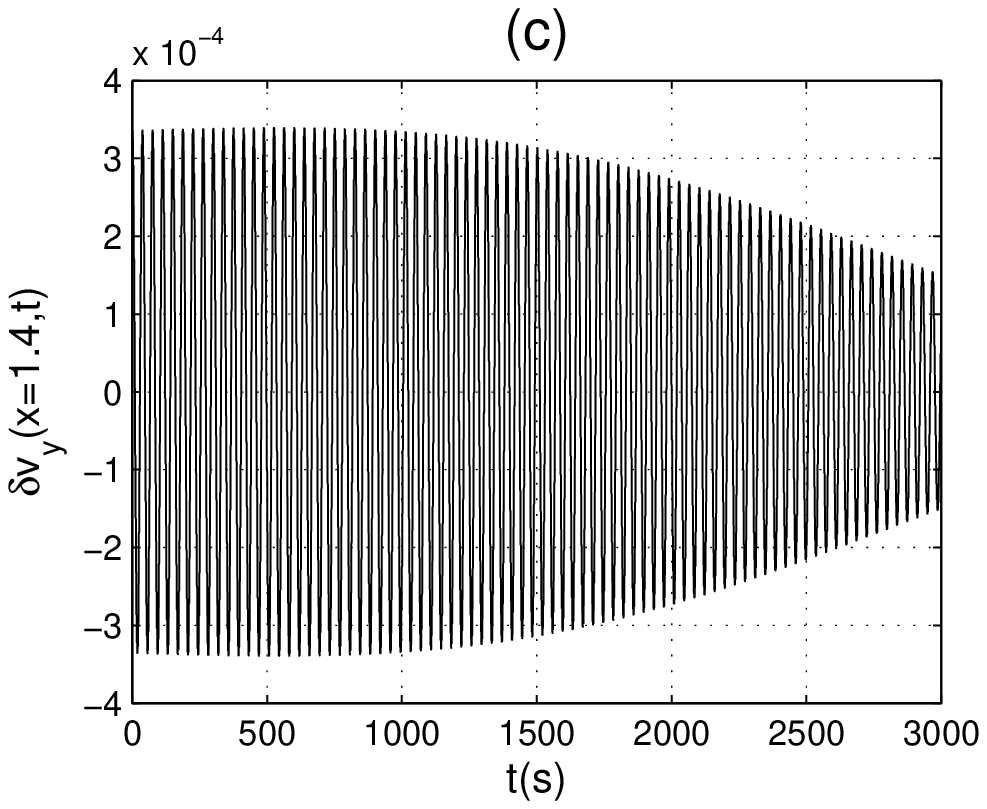} \vspace{2 cm} \caption[]{Same
as Fig. \ref{setA-v} for $R=S=10^8$.} \label{setB-v}
\end{figure}

\begin{figure}
\includegraphics{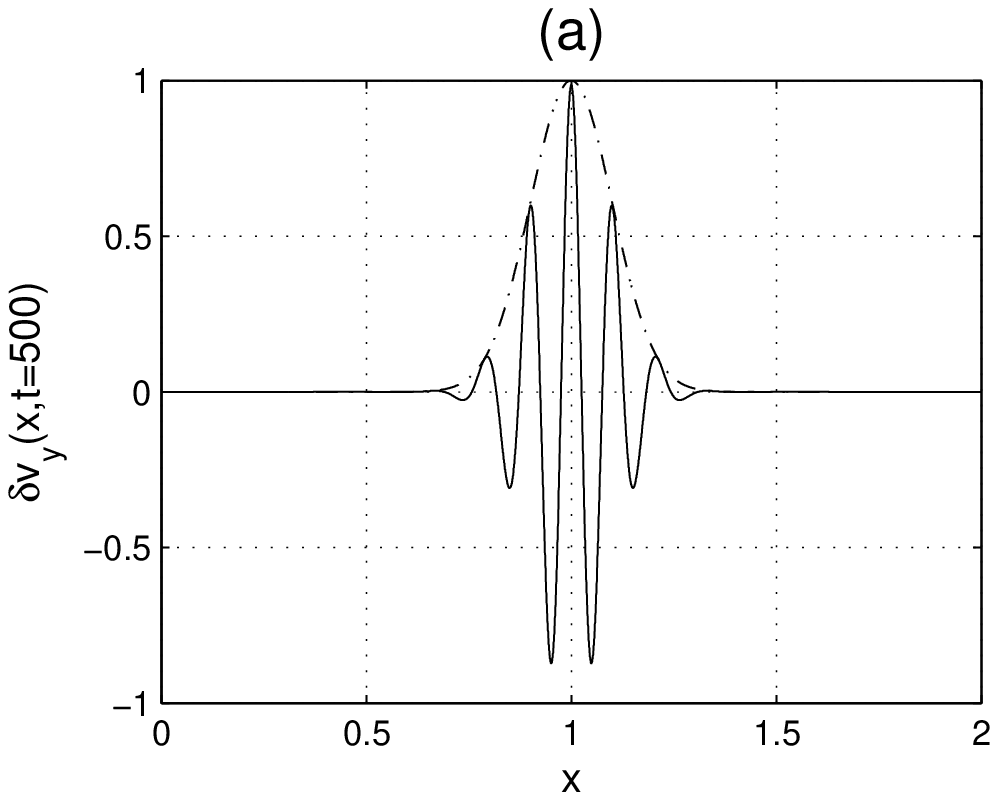} \includegraphics{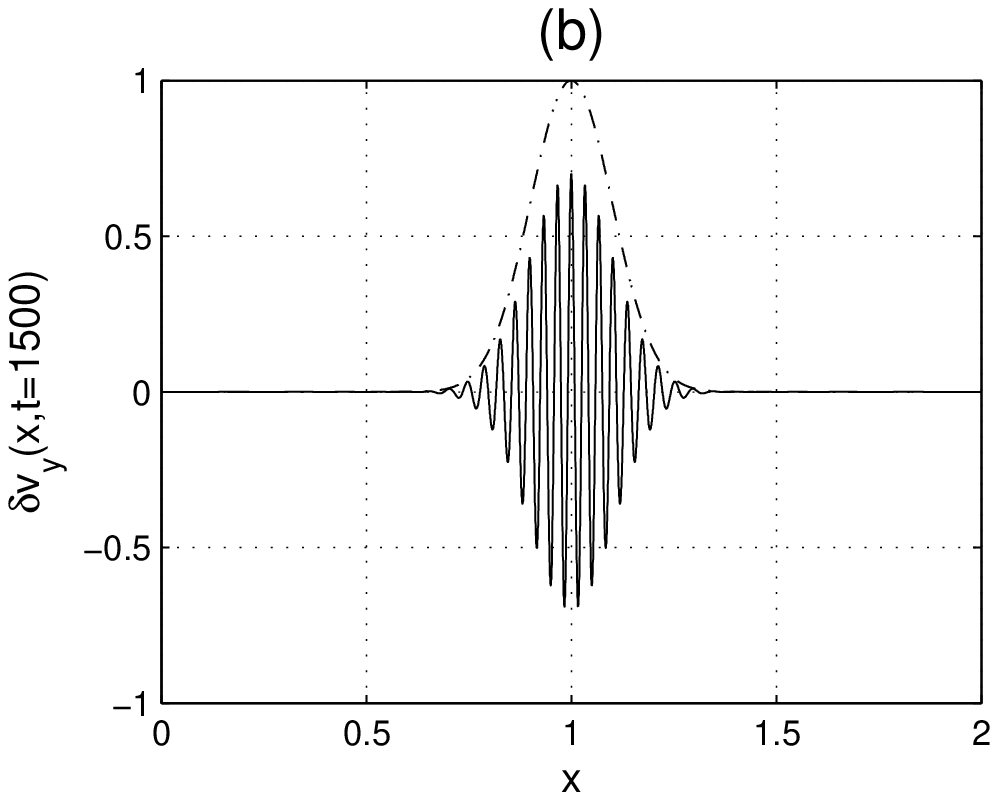} \includegraphics{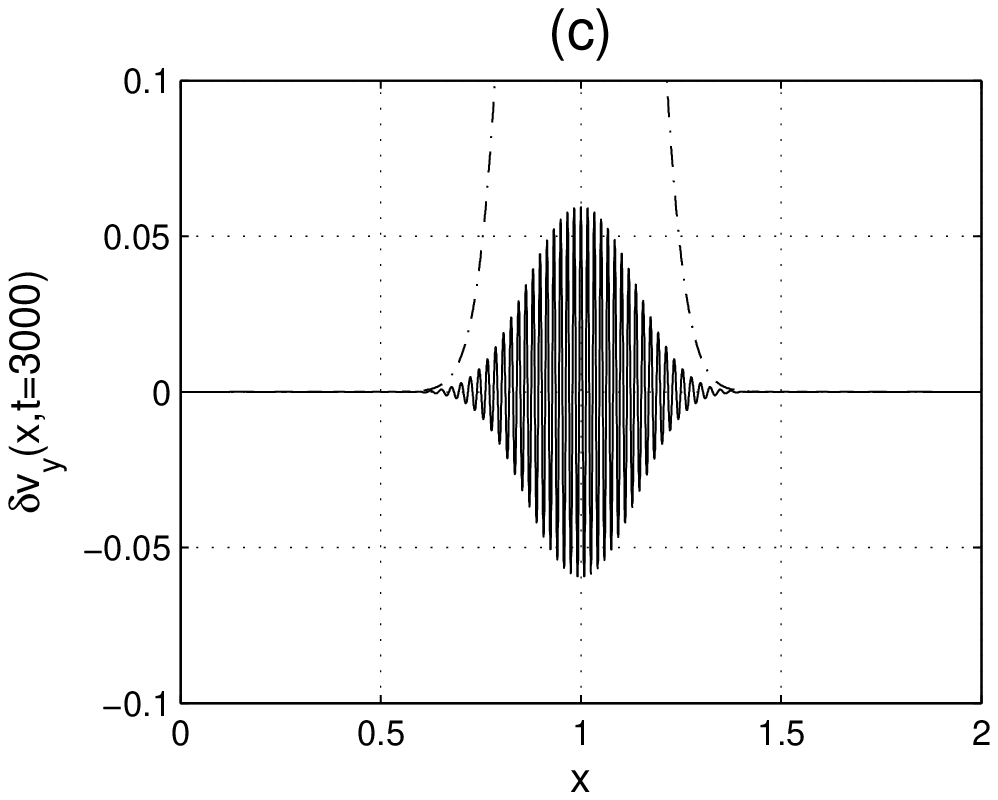} \vspace{2 cm} \caption[]{Same
as Fig. \ref{setA-t} for $R=S=10^8$ at (a) $t=500 s$ ; (b) $t=1500
s$ ; (c) $t=3000 s$.} \label{setB-t}
\end{figure}

\begin{figure}
\includegraphics{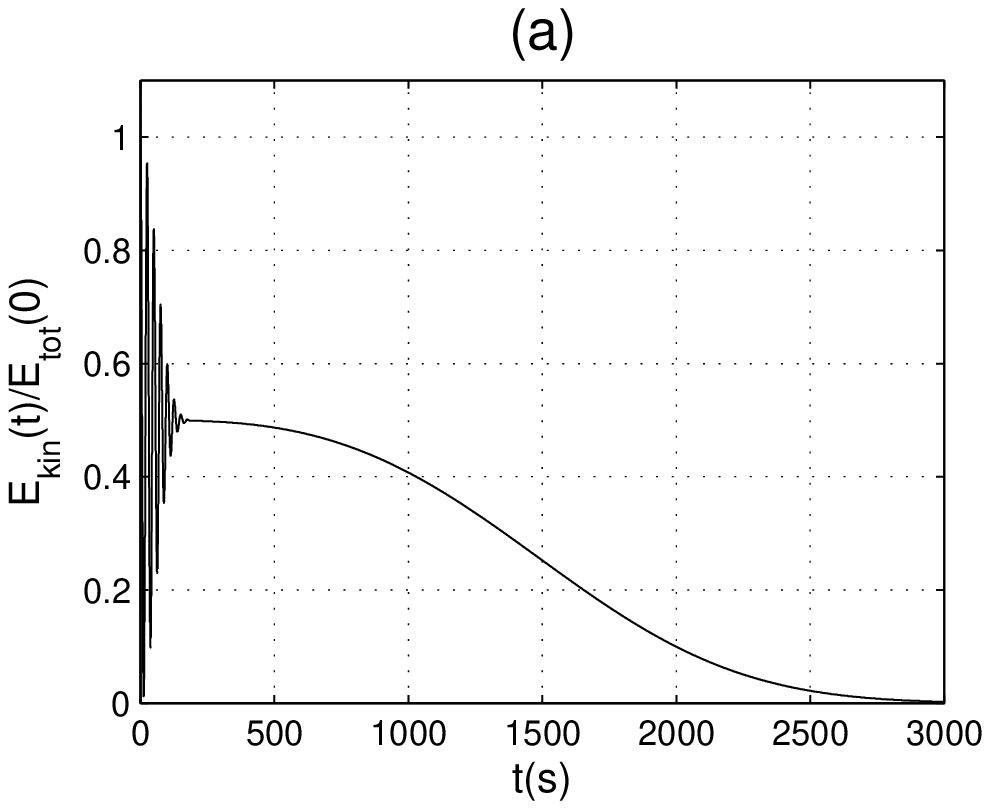} \includegraphics{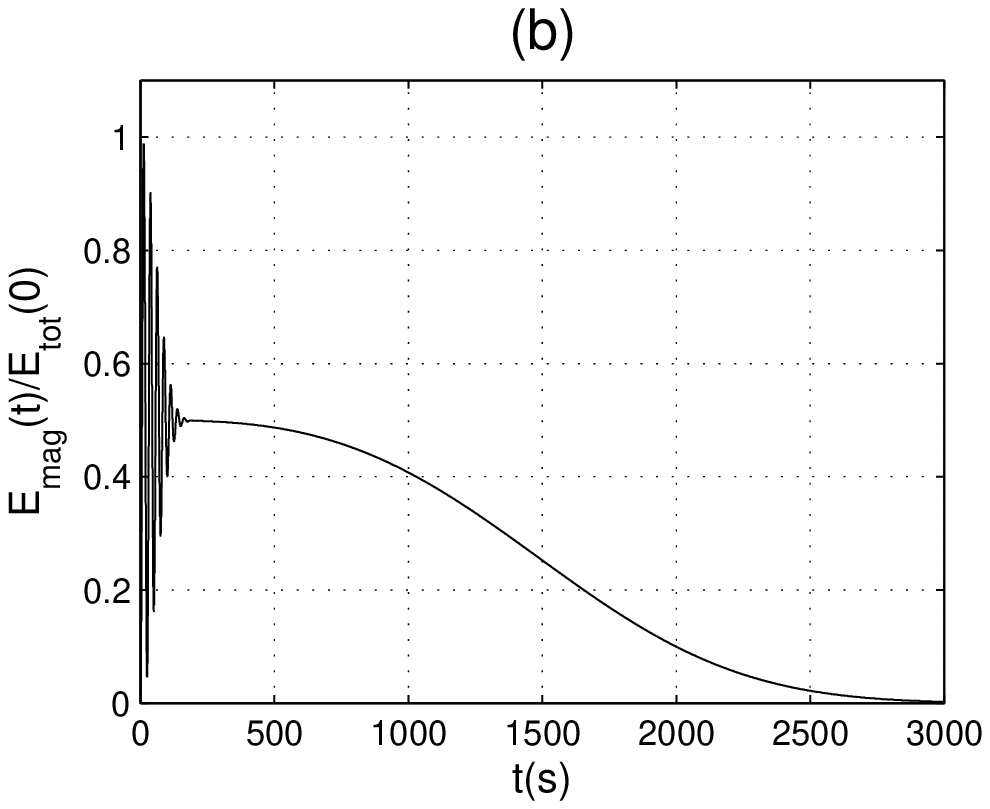} \includegraphics{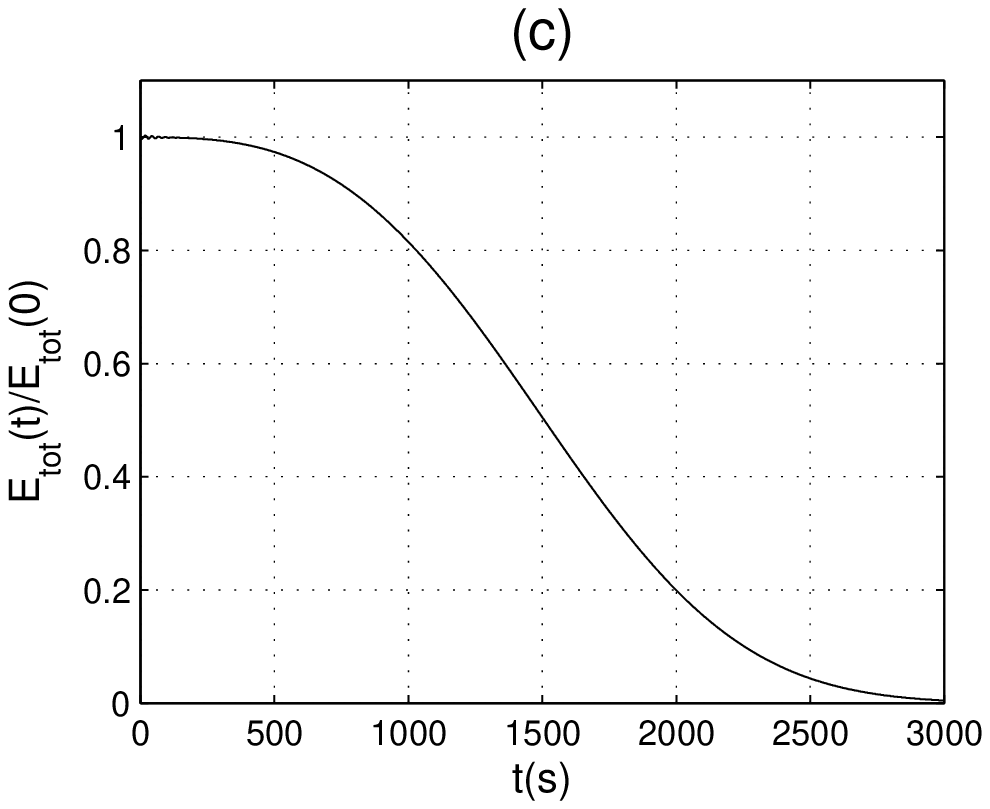} \vspace{2 cm} \caption[]{Same
as Fig. \ref{setA-energy}, for $R=S=10^8$.} \label{setB-energy}
\end{figure}

Figs. \ref{setB-v}-\ref{setB-energy} show evolution of the packet
for the case (b) with $\alpha=2$ and $d=0.1$. Figs.
\ref{setB-v}-\ref{setB-energy} in comparing with Figs.
\ref{setA-v}-\ref{setA-energy} present that in high Reynolds and
Lundquist numbers, i.e. weak damping, the wave packet is damped in
developed stage of phase mixing and it's defocousing is
negligible.

From Eq. (\ref{Energy2}) for $R=S=10^4$, $\alpha=2$ and $d=0.1$,
we obtain $\tau_{\rm dam}=79.1 ~\rm s$. From Eq. (\ref{Pav}) for
$\alpha=2$, the average period of the fundamental mode,
$k=\frac{\pi}{100}$, is obtained as $P_{\rm
avg}^{\alpha=2}=57.9~\rm s$. Therefore the ratio of the damping
time to the average period, $\tau_{\rm dam}/P_{\rm
avg}^{\alpha=2}$ for the fundamental mode is 1.4. From Eq.
(\ref{Energy2}) for $R=S=10^8$, $\alpha=2$ and $d=0.1$, $\tau_{\rm
dam}=1702.6~ \rm s$ and $\tau_{\rm dam}/P_{\rm
avg}^{\alpha=2}\simeq 29.4$. For $R=S=10^8$, if we set $\alpha=4$
and $d=0.05$ then $P_{\rm avg}^{\alpha=4}=62.1\rm~s$, $\tau_{\rm
dam}=1072.7\rm~s$ and $\tau_{\rm dam}/P_{\rm avg}^{\alpha=4}\simeq
17.3$. These strong damping times are in agreement with the
results observed by Nakariakov et al. (1999) and Wang \& Solanki
(2004) deduced from the observation of TRACE.

To test the validity of HP83's damping law in the both weak
damping and strong phase mixing limit, we fit the functional form
$\exp(-At^B )$ on the envelope of $\delta v_y(x,t)$ at $x=1$. Note
that $B=3$ and $A(x)=\frac{1}{6}\nu k^2 v_{\rm A}^ {'2}(x)$ in
HP83, where $\nu=av_{\rm A_0}(R^{-1}+S^{-1})$. The numerical
results obtained for $A$ and $B$ for different ranges of the
Reynolds and Lundquist numbers are tabulated in Table
\ref{tabelA-B}. It shows that for $R=S=10^4-10^{10}$, the
numerical values of $B$ converge to its analytical value but there
is one to four order of magnitude difference between the numerical
and analytical values of $A$. This returns to keeping the term
$T(x,t)$ in Eq. (\ref{main}) which has been missed in HP83. Table
\ref{tabelA-B} also shows that for $S=10^{12}$ and
$R=10^{7}-10^{9}$, the contribution of $T(x,t)$ becomes negligible
in Eq. (\ref{main}) and the numerical values of $A$ and $B$
converge to their corresponding analytical values in HP83's
damping law. Finally one can conclude that the exponential damping
law in time of HP83, $\exp(-At^3)$, is valid for $(R,S)\geq 10^7$.
%-----------------------------------------------------------------------------------------------
\section{Conclusions}
Phase mixing of a packet of standing Alfv\'{e}nic pulses in
fundamental mode is studied. Using a finite difference method, the
linearized MHD equations for a zero-$\beta$ plasma are solved,
numerically. The damping times of oscillations in presence of the
both viscous and resistive dissipations are calculated,
numerically. They are in good agreement with the TRACE
observations. The exponential damping law in time of HP83,
$\exp(-At^3)$, for the different ranges of the Reynolds and the
Lundquist numbers are examined. Our numerical results shows that
it is valid for $(R,S)\geq 10^7$.
%-----------------------------------------------------------------------------------------------
\\
\noindent{{\bf \\Acknowledgements}}. This work has been supported
financially by Research Institute for Astronomy $\&$ Astrophysics
of Maragha (RIAAM), Maragha, Iran.
%-----------------------------------------------------------------------------------------------

%-----------------------------------------------------------------------------------------------
\clearpage
%-----------------------------------------------------------------------
%-----------------------------------------------------------------------
\clearpage
%-----------------------------------------------------------------------
%-----------------------------------------------------------------------
%-----------------------------------------------------------------------
\clearpage
%-----------------------------------------------------------------------
%-----------------------------------------------------------------------
%-----------------------------------------------------------------------
\clearpage
\begin{table}
\centering\caption[]{Numerical and analytical (based on HP83)
values of $A$ and $B$ for some Reynolds and Lundquist numbers in
the cases of $S=R$ and $S\gg R$. Note that $B_{\rm analytic}=3$ in
HP83.}

\begin{tabular}{lllll}\hline\noalign{\smallskip}

 R &S &$A_{\rm numeric}$  &$A_{\rm analytic}$  & $B_{\rm numeric}$ \\\hline\noalign{\smallskip}

\hline\noalign{\smallskip}
$10^4$ &$10^4$& $2.53\times 10^{-3}$ & $2.63\times 10^{-7}$ &$1.277$   \\
$10^5$ &$10^5$& $1.73\times 10^{-5}$ & $2.63\times 10^{-8}$ &$2.001$   \\
$10^6$ &$10^6$& $2.69\times 10^{-8}$ & $2.63\times 10^{-9}$ &$2.692$   \\
$10^7$ &$10^7$& $1.89\times 10^{-9}$ & $2.63\times 10^{-10}$ &$2.914$   \\
$10^8$ &$10^8$& $1.23\times 10^{-10}$ & $2.63\times 10^{-11}$ &$2.980$   \\
$10^9$ &$10^9$& $1.08\times 10^{-11}$ & $2.63\times 10^{-12}$ &$2.992$   \\
$10^{10}$ &$10^{10}$& $1.05\times 10^{-12}$ & $2.63\times 10^{-13}$ &$2.995$   \\
\hline\noalign{\smallskip} \hline\noalign{\smallskip}
$10^7$ &$10^{12}$& $7.70\times 10^{-10}$ & $1.32\times 10^{-10}$ &$2.946$   \\
$10^8$ &$10^{12}$& $5.84\times 10^{-11}$ & $1.32\times 10^{-11}$ &$2.987$   \\
$10^9$ &$10^{12}$& $5.36\times 10^{-12}$ & $1.32\times 10^{-12}$ &$2.993$   \\
\hline\noalign{\smallskip}
\end{tabular}\\
\label{tabelA-B}
\end{table}
%----------------------------------------------------
%\end{multicols}

\end{document}